\documentclass{aa}

\usepackage{epsfig}
\usepackage{times}

\newcommand{\cps}{\mbox{$\,\rm{cts}\,\rm{s}^{-1}$}}
\newcommand{\ergscms}{\mbox{$\,\rm{ergs}\,\rm{cm}^{-2}\,\rm{s}^{-1}$}}
\newcommand{\Grad}{\mbox{$\,^{\circ}$}}
\newcommand{\NH}{\mbox{$\rm{N}_{\rm{H}}$}}
\newcommand{\kms}{\mbox{$\,\rm{km}\,\rm{s}^{-1}$}}
\newcommand{\alphaox}{\mbox{$\alpha_{\rm{OX}}$}}

\begin{document}

\thesaurus{03 (11.01.2; 11.02.1; 13.25.2)}

\title{On the evolutionary behaviour  of BL\,Lac  objects\thanks{Based 
    on observations from the German-Spanish Astronomical 
        Center, Calar Alto, operated 
        jointly by the Max-Planck-Institut f\"ur
        Astronomie, Heidelberg, 
        and the Spanish National 
        Commission for Astronomy, 
        and from the William Herschel Telescope operated on the
        island of La Palma by the Royal Greenwich Observatory 
        in the Spanish Observatorio del Roque de los Muchachos of the
        Instituto de Astrof\'{i}sica de Canarias}
}

\author{N. Bade\inst{1}  \and V. Beckmann\inst{1} 
  \and N.G. Douglas\inst{2} \and P. D. Barthel\inst{2} \and 
  D. Engels\inst{1} \and L.
  Cordis\inst{1} \and P. Nass\inst{3} \and W. Voges\inst{3}}
 
\offprints{N. Bade; e--mail: nbade@hs.uni-hamburg.de}
 
\institute{Hamburger Sternwarte, Gojenbergsweg 112, 
           D-21029 Hamburg, Germany 
           \and
           Kapteyn Astronomical Institute, P.O.Box 800, 9700 AV Groningen,
           The Netherlands       
           \and 
           MPI f\"ur Extraterrestrische Physik,
           D-85740 Garching, Germany 
}

\date{Received date; accepted date} 
 
\authorrunning{Bade et al.} 

\maketitle

\sloppy
 
\begin{abstract}
We present a new well defined sample of BL\,Lac objects selected from the
ROSAT All-Sky Survey (RASS). The sample consists of 39 objects with 35 forming
a flux limited sample down to 
$\rm f_{X}(0.5 - 2.0\,keV) =  
8\cdot 10^{-13}\,ergs\,cm^{-2}\,s^{-1}$, 
redshifts
are known for 33 objects (and 31 of the complete sample). 
X-ray spectral properties were determined
for each object individually with the RASS data. The luminosity function of
RASS selected BL\,Lac objects is compatible with results provided by objects
selected with the {\em Einstein} observatory, but the RASS selected sample
contains objects with luminosities at least tenfold higher. 
Our analysis confirms the negative evolution for X-ray selected BL\,Lac
objects found in a sample by the {\em Einstein} observatory, the
parameterization provides similar results.
A subdivision of
the sample into halves according to the X-ray to optical flux ratio
yielded unexpected results. The extremely X-ray dominated objects have higher
redshifts and X-ray luminosities and only this subgroup shows clear signs of
strong negative evolution. The evolutionary behaviour of objects with an
intermediate spectral energy distribution between X-ray and radio dominated
is compatible with no evolution at all. 
Consequences for unified
schemes of X-ray and radio selected BL\,Lac objects are discussed.
We
suggest that the intermediate BL\,Lac objects are the basic BL\,Lac population.
The distinction between the two subgroups can be explained if extreme X-ray
dominated BL\,Lac objects are observed in a state of enhanced X-ray activity.


\end{abstract}

\section{Introduction}
The most common view about BL\,Lac Objects is that we are looking into a highly
relativistic jet (Blandford \& Rees, \cite{blandford}).
The high variability, the
polarization, and the spectral energy distribution can 
 %NGD% 
 in principle all
be explained with this
model. 
But there are still unsolved problems for this model. Examples are the nature
of the mechanism(s) that generates and collimates the jet. 
The evolution of physical parameters as the energy and momentum
density along the jet is not yet clear, too. 
An important question is also whether
the jets are composed of highly relativistic hadronic or leptonic plasma
(Kollgaard, \cite{kollgaard}).
Furthermore there exists
the competing 
 %NGD%  model 
theory that at least some BL\,Lac objects 
originate from microlensed QSO (Ostriker \& Vietri, \cite{ostriker}).
Models of BL\,Lac objects must explain why X-ray and radio selected
BL\,Lac objects differ in some 
observational characteristics (degree of variability and polarization
(Jannuzi et al., \cite{jannuzi}))
so that it is not
necessarily true that both object classes 
are of the same astrophysical origin.
A sensitive test for the models is the cosmological
evolutionary behaviour.

To evaluate the luminosity function and evolution behaviour of an object
class complete samples with known distances of each individual object
are necessary.  For BL\,Lac objects two problems exist which render the
determination of their luminosity function difficult.
First, on account of the inconspicuous optical spectral properties of 
 BL\,Lacs (the absence of strong
 emission and absorption lines being one of the defining criteria)
 more effort is required in their selection than is the case with
 most other object classes. 
Although there are defining criteria suitable for
optical selection (variability, polarisation) no optically selected
sample of BL\,Lac objects with more than 10 objects 
exists (Kollgaard, \cite{kollgaard}).
Second, their redshifts are difficult to determine for the same reason.

As a consequence, flux limited samples of BL\,Lac objects with nearly complete
redshift information are rare. Stickel et al. (\cite{stickel}) 
selected a
sample of 35 objects in the radio wavelength region with radio fluxes above
1\,Jy. They found a flat
redshift distribution between $0.1 < \rm{z} < 1.2$ and their analysis of the
sample revealed positive evolution; radio selected BL\,Lac objects are
more numerous in cosmological distances than in our local neighbourhood. 
At  X-ray wavelengths 
a complete well defined sample was built up with the EMSS (Stocke
et al., \cite{emss}). Morris et al. (\cite{morris}) presented a sample of 22
objects down to 
$f_{X}(\rm 0.3 - 3.5\,keV) = 5\cdot 10^{-13}\,ergs\,cm^{-2}\,s^{-1}$, which
was later expanded to 30 objects and lower fluxes by Wolter et
al. (\cite{wolter}). The EMSS objects exhibit a redshift
distribution with
strong concentration to low redshifts $\rm z < 0.3$ and the sample shows clear
signs for negative evolution. The EMSS sample was constructed with 
serendipitously found BL\,Lac objects from pointed observations of the 
{\em Einstein} satellite.
Because most these pointed observations 
were of short exposure,
objects with low
fluxes are less frequent and are rare in the sample. 

We contribute to this discussion with a new
sample consisting of 39 X-ray selected BL\,Lac objects 
with redshifts available for more than $80\%$ of them. 
The selection is 
based on the ROSAT All-Sky Survey (RASS, Voges et al., \cite{rass}).
With a flux
limit of $f_{X}(\rm 0.5 - 2.0\,keV) = 8\cdot 10^{-13}\,ergs\,cm^{-2}\,s^{-1}$ 
and redshifts up to $\rm z \sim 0.8$ the sample is suitable to test the
evolutionary behaviour of the sample and to determine the luminosity function
of RASS selected BL\,Lac objects. 
In an earlier paper (Nass et al., \cite{nass}) we presented a larger sample to
discuss selection processes of BL\,Lac objects within the RASS, but we could
not deal with the evolutionary behaviour because the flux limit was not deep
enough and many redshifts were unknown. The new sample consists of objects 
with a wide range of spectral energy distributions, from extremely X-ray
dominated objects to objects with intermediate spectral energy distribution.
Objects with a maximum of $\nu f_{\nu}$ in the radio or mm wavelength region,
which are frequently found among the 1\,Jy sample, are not contained in our
X-ray flux limited sample.
We were able to subdivide the
sample into two subgroups according to \alphaox\footnote{we define \alphaox\ 
  as the power law index between 1\,keV and 4400\,\AA\ with $\rm f_{\nu}
  \propto \nu^{-\alphaox}$}. 
Several properties are different in these subgroups. 

The following section describes the observational characteristics of the 
sample 
in the radio-, optical- and X-ray region.
Details of the redshift determination are discussed, bcause it is a delicate 
process for BL\,Lac objects. 
Section III analyses the properties of the sample, in particular the
luminosity function and the evolutionary behaviour.
In Sect. IV the new results  are discussed in the context of unified schemes
for BL\,Lac objects.
Several new objects need comments to their redshift
determination which are presented in an appendix.

We use cosmological parameters $\rm H_{0} =
50\,km\,s^{-1}\,Mpc^{-1}$ and $q_{0} = 0$ in this paper. 

\section{Observational basis}
\subsection{Sample and object definition \label{sampledefinition}}
The sample of BL\,Lac objects presented here
is a subset of the Hamburg ROSAT
X-ray bright sample (HRX, Cordis et al., \cite{hrx}) of AGN. The HRX was 
pre-identified on direct and objective prism plates which were taken with the
Hamburg Schmidt telescope on Calar Alto\,/\,Spain within the Hamburg Quasar
Survey (HQS, Hagen et al., \cite{hqs}).
Obvious galactic counterparts were excluded from 
further study (Bade et al., \cite{hrc}). 
Slit spectroscopy was performed in order to obtain reliable
classifications of the sources unidentified on the Schmidt plates and to 
determine redshifts for the
AGN candidates. 
Galaxy clusters are not subject of the HRX and they have not been
investigated in detail up to now.

The sample  is drawn from two subareas of the HRX.
The area with $ 45\Grad < \delta < 70\Grad$ and $\rm 8h < \alpha < 17h$
covers 1687\,deg$^{2}$ and it is completely identified 
down to a hard ROSAT PSPC (0.5 $-$ 2.0\,keV)
countrate of 0.075\cps (corresponding to $\rm f_{x}(0.5 - 2.0\,keV) = 8\cdot
10^{-13}\ergscms$).  A detailed
description of this core HRX area can be found in an accompanying paper 
(Cordis et al., in preparation) which also provides identifications for 
all X-ray sources in this region.
A second area with $\delta < 45\Grad$ was added for this paper and
covers 1150\,deg$^{2}$.
It is complete down to a hard countrate of 0.15\cps.
Table~\ref{areazwei} presents the boundaries of this patchy area.

\begin{table}
\caption[]{Field boundaries for the surveyed areas with $\delta < 45\Grad$.
They are complete down to a hard ROSAT PSPC countrate of 0.15\cps 
\label{areazwei}}
\begin{center}
\begin{tabular}{llll}
R.A.$_{min}$\rule[-2mm]{0mm}{2mm} & R.A.$_{max}$ & Decl.$_{min}$ & 
Decl.$_{max}$ \\
\hline
08h\,45m\rule{4mm}{0mm} & 09\,h30m\rule{8mm}{0mm} & 
19\Grad$53'$\rule{4mm}{0mm} & 25\Grad$00'$\\
15h\,55m & 16\,h18m & 19\Grad$57'$ & 25\Grad$00'$\\
07h\,33m & 13\,h51m & 25\Grad$00'$ & 30\Grad$00'$\\
16h\,01m & 17\,h33m & 25\Grad$00'$ & 30\Grad$00'$\\
07h\,32m & 13\,h44m & 30\Grad$00'$ & 35\Grad$20'$\\
16h\,02m & 17\,h37m & 30\Grad$00'$ & 35\Grad$14'$\\
16h\,20m & 16\,h49m & 39\Grad$54'$ & 45\Grad$00'$\\
    
\end{tabular}
\end{center}

\end{table}

Table~\ref{basistabel} presents BL\,Lac objects located inside the HRX areas.
For the calculation of the luminosity function
we restricted ourselves to objects with the above given countrate limits.
Their names are marked with an asterisk.
In previous identification projects BL\,Lac objects were more frequent among
the last unidentified X-ray sources. Only one X-ray source remained
unidentified because we did not take a spectrum from this object. Since the
HRX comprises 302 X-ray sources (among them 201 extragalactic sources)
this means a completeness level of $99.7\%$. Selection effects which could be
introduced by our identification process are discussed in Sect. 
\ref{analysis}.

Our criteria to classify X-ray sources as BL\,Lac objects are essentially
spectroscopically similar 
to the work of the EMSS. By definition the spectra of BL\,Lac
objects are dominated by a nonthermal continuum. Explicitly the selection
criteria are: 
\begin{itemize}
\item Emission lines with $\rm
W_{\lambda} > 5\AA$ must be absent.
\item The contrast of the Ca II break from 
the hosting galaxy
must be less than $25\%$ in the optical spectrum.
\end{itemize} 
The gathered radio data supported us in selecting our BL\,Lac candidates, but
we did not use radio data to exclude objects from our list of candidates. 
We did not measure the polarization
and did not check the optical variability behaviour of the newly
presented objects up to now. 
Detailed studies of the EMSS BL\,Lac objects have shown
that X-ray sources with the above mentioned selection criteria are all radio
sources (Stocke et al., \cite{stocke})  and most of them are 
variable and the duty cycle of polarized emission (fraction of time spent
with the degree of polarization greater than $4\%$) is $44\%$ (Jannuzi et al.,
\cite{jannuzi}).

\begin{table*}
\caption[]{BL\,Lac objects with hard ROSAT countrate above $\rm
  0.075\,cts\,s^{-1}$ in the surveyed area. Objects forming the flux limited
  sample are marked with an asterisk. 
 Positions are measured with an
  accuracy of $2\arcsec$ on the HQS direct plates, 
  or are taken from  POSS\,II in the case of
  objects too faint to be visible on the HQS plates.
  Column ``cps'' lists the countrate $[s^{-1}]$ for the hard ROSAT band ($0.5
  -2.0\,\rm{keV}$), 
  ``$\rm f_{X}$'' is
  the X-ray flux
  $\rm [10^{-12}\,ergs\,cm^{-2}\,s^{-1}]$, 
  ``mag'' is the optical magnitude B,
  ``$\rm f_{R}$'' is the radio flux [mJy] from the NVSS at 1.4\,Ghz, 
  ``$\Gamma$'' is the
  photon index ($\rm N(E) \propto E^{-\Gamma})$), 
  ``$\rm \log L_{X}$'' is the 
  logarithm of the 
  monochromatic
  X-ray luminosity $\rm [W\,Hz^{-1}]$ at 2\,keV, 
  ``$\rm \log L_{R}$'' is  the 
  logarithm of the
  monochromatic
  radio luminosity $\rm [W\,Hz^{-1}]$ at the appropriate frequency,
``\alphaox '' is the optical-xray spectral index as defined in
 footnote 1 and
 ``$\alpha_{R}$'' is the radio spectral index between 1.4 and 4.85\,GHz
 \label{basistabel}}
\begin{tabular}{lrrlrrrrrrrrrr}
Name & \multicolumn{2}{c}{R.A. (2000.0)~~  Decl.} & z & cps & $\rm f_X$ & 
mag \rule[-2mm]{0mm}{2mm} & 
$\rm f_R$ & $\Gamma$ & { \tiny $\rm \log L_{X}$} & $\rm M_{abs}$ & 
{ \tiny $\rm \log L_{R}$} & $\alpha_{OX}$  &  $\alpha_{R}$ \\
\hline
RX\,J0809.8+5218$^{*}$  & 08 09 49.0  &  52 18 56 &  0.138 &  0.371  &  4.68 &  15.6 &
183 &  2.96 &   19.54 &   $-$24.1 & 25.0 & 1.36 &  $-$0.3\\
RX\,J0916.8+5238$^{*}$  & 09 16 52.0  &  52 38 27  & 0.190  & 0.175  &  2.00  & 19.7
&  83  & 2.18  &  19.70 & $-$20.7 & 25.0 &   0.82 & $-$0.5\\
RX\,J0930.6+4950$^{*}$  & 09 30 37.6   & 49 50 24  & 0.186  & 1.154  & 13.45 &  18.0
& 22  & 1.88  &  20.58  & $-$22.4 & 24.5 &  0.74 &  $-$0.3\\
RX\,J1008.1+4705$^{*}$ &  10 08 11.4 &   47 05 20 &  0.343 &  0.383  &  4.32 &  18.9
&   5 &  2.14  &  20.63  & $-$22.9 & 24.6 &  0.81 &  $-$0.1\\
GB1011+496$^{*}$   &    10 15 04.0  &  49 25 59 &  0.200  & 0.594 &   6.64  & 16.5 &
378  & 2.26  &  20.26 &  $-$24.0 & 25.7 & 1.10 &  $-$0.4\\
RX\,J1031.3+5053$^{*}$ &  10 31 18.6  &  50 53 34  & 0.361  & 1.561 &  17.75 &  16.8 &
38 &  2.31  &  21.26 & $-$25.1 & 25.3 &  0.91 &  $-$0.4 \\
RX\,J1058.6+5628$^{*}$ &  10 58 37.8  &  56 28 09  & 0.144 &  0.120  &  1.33  & 15.8 &
229  & 2.32 &   19.23  & $-$24.0 & 25.3 & 1.45 &  $-$0.2 \\
Mrk180$^{*}$      &     11 36 26.6  &  70 09 25 &  0.046 &  1.711 &  19.57  & 14.7 &
222  & 2.26 &   19.37   & $-$22.5 & 24.4 & 1.18 &  $-$0.4 \\
RX\,J1136.5+6737$^{*}$ &  11 36 30.3  &  67 37 05  & 0.135  & 0.977  & 11.40  & 17.6
&  46  & 1.89  &  20.21  & $-$22.0 & 24.5 & 0.82 &  $-$0.1\\
MS\,12292+6430$^{*}$  & 12 31 31.5   & 64 14 16  & 0.164 &  0.166  &  1.96 & 
17.4 & 59  & 2.02  & 19.59 & $-$22.7 & 24.8 & 1.14 & 0.0\\
MS\,12354+6315$^{*}$  & 12 37 39.1   & 62 58 41  & 0.297 &  0.139  &  1.64 & 
19.0 & 13  & 1.89  & 20.11 & $-$22.5 & 24.7 & 0.92 & \\
RX\,J1248.3+5820$^{*}$  & 12 48 18.5   & 58 20 27  &      &  0.152  &  1.70 &  16.3
& 245  & 2.42  &        & & & 1.35 & $-$0.1\\
RX\,J1302.9+5056$^{*}$  & 13 02 55.5  &  50 56 17 &  0.688 &  0.241 &   2.78  & 20.2
&   3  & 1.94 &   21.19 & $-$22.9 & 24.8 &  0.60  &  0.8 \\
RX\,J1324.0+5739$^{*}$ &  13 24 00.0  &  57 39 16 &  0.115  & 0.096 &   1.11 &  17.3
&  40  & 2.04  &  19.01  & $-$21.5 & 24.4 & 1.16 &  $-$0.4\\
RX\,J1353.4+5601$^{*}$  & 13 53 28.0  &  56 00 55 &  0.370  & 0.114  &  1.31 &  18.9
&  15 &  2.01 &   19.96  & $-$22.6 & 24.8 & 0.91 &  $-$0.2\\
RX\,J1404.8+6554$^{*}$  & 14 04 49.6  &  65 54 30 &  0.364 &  0.091  &  1.06  & 19.0
&  15 &  2.26  &  20.06 & $-$22.5 & 25.0 & 0.95  &  \\
RX\,J1410.5+6100$^{*}$ &  14 10 31.7 &   61 00 10 & 0.384 &  0.116  &  1.36 &  20.1  &
12 &  1.90 &   20.28 &  $-$21.5 & 24.4 & 0.72  &  $-$0.8\\
RX\,J1422.6+5801$^{*}$ &  14 22 39.0 &   58 01 55 &  0.638 &  0.867 &   9.99 &  18.7
&  13  & 2.06 &   21.65 & $-$24.1 & 24.9 & 0.63 &  $-$0.8\\
RX\,J1436.9+5639$^{*}$ &  14 36 57.8  &  56 39 25 &       & 0.087  &  1.00 &  18.8 &
21 &  2.22 &        & & &  1.06 &  $-$0.5\\
1E14435+6349 $^{*}$ &   14 44 36.6  &  63 36 26 &  0.299 &  0.091 &   1.07 &  19.6
&  19 &  1.83 &   19.94  & $-$21.4 & 24.9 & 0.82 & \\
RX\,J1451.4+6354$^{*}$ &  14 51 27.5 &   63 54 19 &  0.650  & 0.077 &   0.89  & 19.6
&     &  2.38  &  20.59   & $-$23.3 & & 0.91  &    \\
RX\,J1456.0+5048$^{*}$  & 14 56 03.7  &  50 48 25 &  0.480 &  0.734  &  8.58 &  19.3
&   4 &  2.06 &   21.28  & $-$23.3 & 24.7 & 0.64 &  \\
RX\,J1458.4+4832$^{*}$ &  14 58 28.0  &  48 32 40 &  0.539 &  0.224 &   2.68 &  20.2
&  3 &  1.90 &   20.92 & $-$22.2 & 24.6 &  0.61 &  \\
RX\,J1517.7+6525$^{*}$ &  15 17 47.6 &   65 25 21  &      & 0.673 &   8.01 &  15.5
&  39 &  2.06  &      & & &   1.19 &  $-$0.6\\
1ES\,1533+535$^{*}$  &    15 35 00.8  &  53 20 35 &  0.890 &  0.691  &  8.01  & 18.2 &
18 &  1.94   &    21.93  & $-$25.5 & 25.6 & 0.72 &  $-$0.4\\
RX\,J16443+4546$^{*}$  &  16 44 19.8  &  45 46 45 &  0.220 &  0.077  &  0.89 &  18.7 &
170 &  2.36 &   19.45 & $-$22.1 & 25.4 &  1.11 &  $-$0.8\\
IZw187$^{*}$      &     17 28 18.4 &   50 13 10 &  0.055  & 1.212  & 14.59  & 16.0
& 168$^{1}$  & 2.02  &  19.47 & $-$21.1 & 24.3 &  0.95 & \\
RX\,J0809.6+3455$^{*}$  & 08 09 38.5  &  34 55 37  & 0.082 &  0.204  &  2.63  & 17.0
&     224  & 2.87  &  18.83 & $-$21.5 & 24.7 &  0.89 &  0.0\\
RX\,J0832.8+3300 &  08 32 52.0  &  33 00 11  & 0.671 &  0.099  &  1.27  & 20.7 &
4  & 1.57 &   20.84 & $-$22.3 & 24.8 & 0.62 &  0.0\\
B2 0906+31$^{*}$    &   09 09 53.3  &  31 06 02  & 0.274  & 0.185  &  2.18  & 17.8
&    78$^{1}$  & 2.21  &  20.09 & $-$23.0 & 25.5 &  1.01 &  0.2\\
B2\,0912+29$^{*}$ &  09 15 52.2 &   29 33 20  &      & 0.286 &   3.35  & 16.3 &
342 &  2.19  &       & & & 1.15 &  $-$0.5\\
RX\,J1111.5+3452$^{*}$ &  11 11 30.9  &  34 52 01 &  0.212 &  0.233  &  2.73 &  19.5
&    4$^{1}$  & 2.29   & 19.92  & $-$21.2 & 23.8 & 0.81 &  $-$0.3\\
B2 1215+30$^{*}$  &     12 17 52.0  &  30 07 02 &  0.130 &  1.007 &  11.55 &  15.6 &
430 &  2.46  &  20.03  & $-$24.0 & 25.5 &  1.16  & $-$0.1\\
PG1218+304$^{*}$    &   12 21 21.8 &   30 10 37  & 0.182 &  0.776 &   9.05 &  17.7  &
72  & 2.09 &   20.34  & $-$22.6 & 24.9 &  0.86  & 0.0\\
RX\,J1230.2+2518 &  12 30 14.0 &   25 18 07  &      & 0.115 &   1.33 &  15.7 &
351$^{1}$  & 2.14   &     &  & &  1.44  & \\
RX\,J1237.0+3020$^{*}$ &  12 37 05.7  &  30 20 03 &  0.700  & 0.276  &  3.22  & 20.0
&   6 &  1.85  &  21.28  & $-$23.1 & 24.8 &  0.60 &  0.0\\
RX\,J1241.6+3440  & 12 41 41.4  &  34 40 31  &      & 0.091   & 1.06  & 20.2 &
10  & 1.93  &       & & &  0.75 &  0.1\\
RX\,J1326.2+2933 &  13 26 15.0  &  29 33 30  & 0.431  & 0.128  &  1.46  & 19.6
&  28 &  2.11 &   20.39 & $-$22.3 & 24.8 &   0.81  & $-$0.8\\
RX\,J1631.4+4217$^{*}$  & 16 31 24.7 &   42 17 02  & 0.468  & 0.250 &   2.88 &  19.5
&  7  & 1.91 &   20.80 & $-$22.6 & 24.7 &  0.70 &  0.0\\
\end{tabular}
$^{1}$ 4.85\,GHz fluxes measured with VLA D configuration

\end{table*}

\begin{figure*} %einspaltig
\epsfig{file=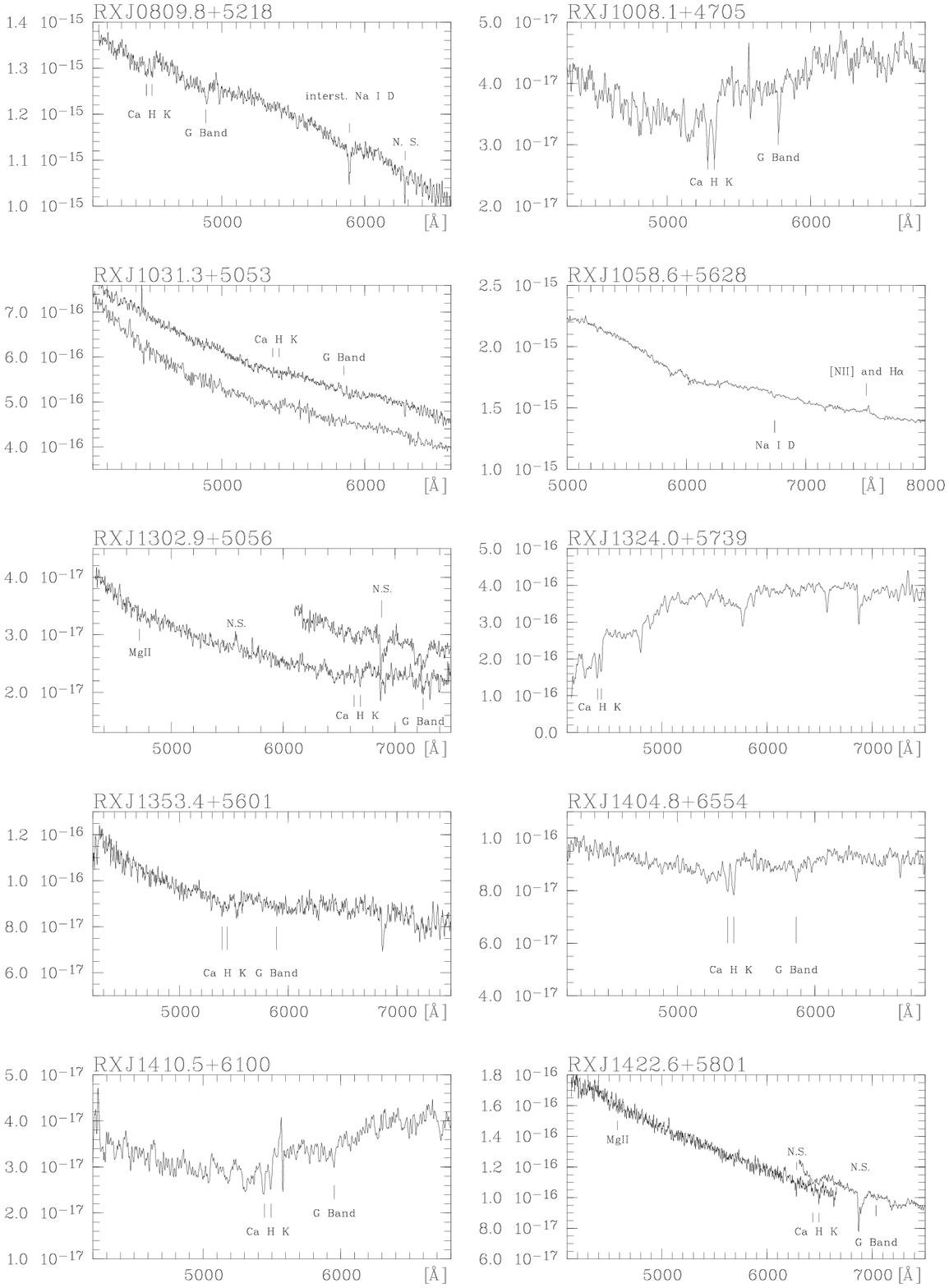,width=18.0cm}
\caption[]{\label{spektren} Optical spectra of objects with newly determined
  redshift. $\rm f_{\lambda}\,[ergs\,cm^{-2}\,s^{-1}]$ is plotted against
  wavelength}
\end{figure*}

\begin{figure*} %einspaltig
\epsfig{file=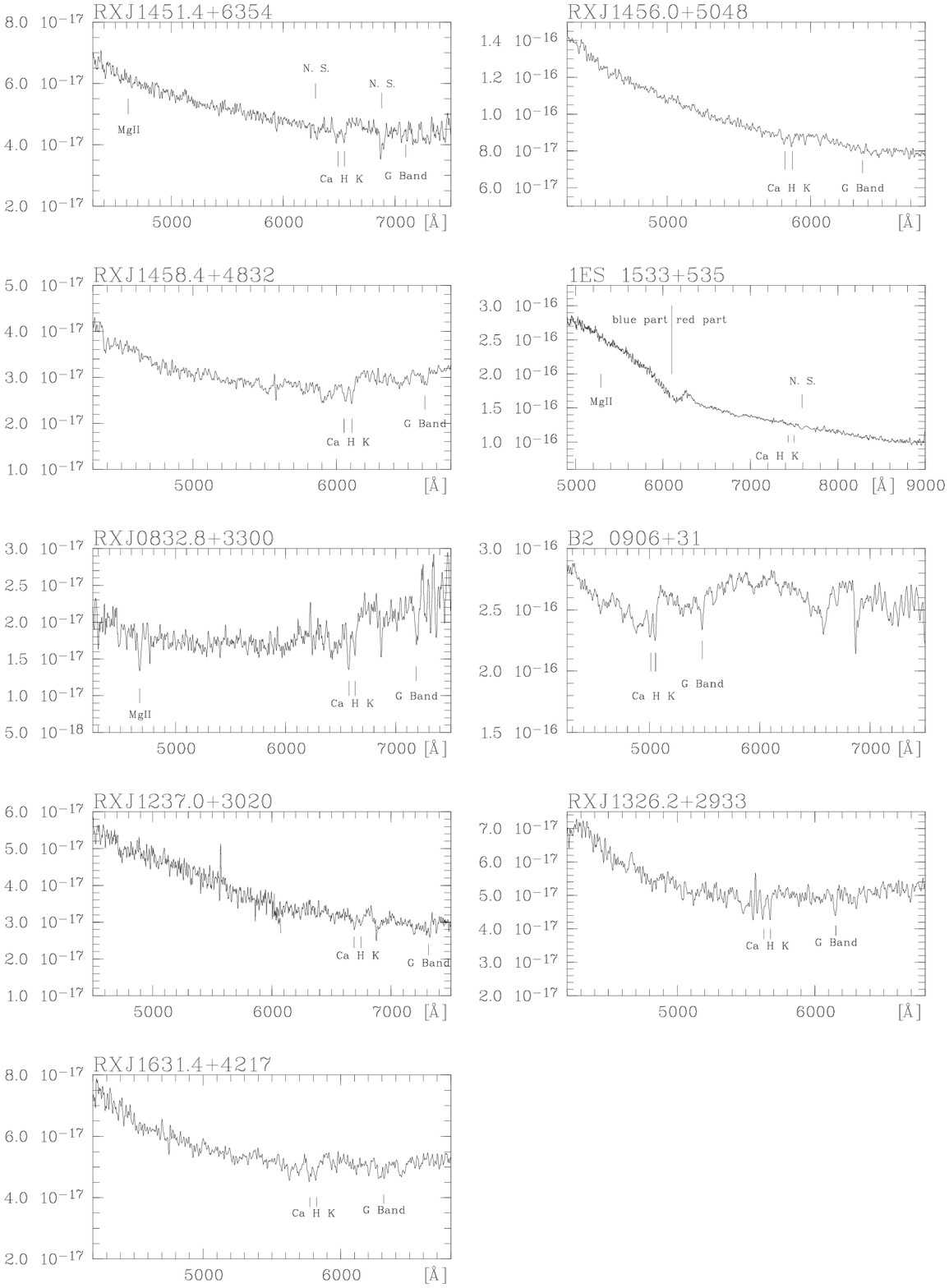,width=18.0cm}
\begin{center}
Continuation of {\bf  Fig. \ref{spektren}}
\end{center}
\end{figure*}

\subsection{Radio properties \label{radio}}
Up to now the presented sample was not observed with dedicated radio
observations. Nevertheless radio properties of BL\,Lac objects are an
important tool to characterize their multifrequency behaviour. Previous own
observations with the VLA at 4.85\,GHz, partly published in Nass et al.,
\cite{nass},  and two
large ongoing VLA surveys at 1.4\,GHz in the northern sky provide
radio fluxes for all but one object. One of the VLA surveys is the NVSS
(Condon et al., \cite{nvss}). For this project the VLA 
is scanning the whole northern sky with $\delta >
-40\degr$ down to 2.5\,mJy with a positional accuracy for point sources
between $1\arcsec$ and $7\farcs 5$. The completion of this project is planned
for the end of 1997
--- at the time of writing of this paper many areas remain
to be observed and processed.

The other survey is the FIRST survey (White
et al., \cite{first}) which is performed with higher resolution and
sensitivity on selected sky fields. 
The patchy second HRX area falls nearly
totally on the FIRST survey area. For all BL\,Lac objects 
in our sample whose position fell within the VLA surveys, a 
radio
counterpart was found.
The only object 
for which we give no radio flux measurement,
RX\,J1451.4+6354, was not covered by own measurements or 
by either of the two VLA surveys. 

For 32 of the 39 objects the two fluxes at 1.4\,GHz and 4.85\,GHz enabled
us to determine the radio spectral index $\alpha_{R}$ ($f_{\nu} \propto
\nu^{\alpha_{R}}$) (see Table~\ref{basistabel}). 
Variability between the two measurements is possible, and
therefore a more reliable determination of the radio spectral index needs
dedicated measurements. There is no $\alpha_{R} < -1$ and most
objects show flat $\alpha_{R} > -0.5$ which is a characteristic property of
BL\,Lac objects (Urry \& Padovani, \cite{urrypadovani}). 
Possibly some of the steeper $\alpha_{R}$
(e.g. $-0.8$ for RX\,J1422.6+5801) are due to variability between the two
measurements.  

\subsection{X-ray properties \label{xray}}
The selection of the objects is based on the ROSAT PSPC countrate between 
0.5\,keV $<$ E $<$ 2.0\,keV in the RASS. 
The conversion factor between flux and countrate
for a power law with low energy absorption by photoionization depends on the
photon index $\Gamma$ and the hydrogen column density \NH.
We assume only Galactic \NH\ taken from the Stark et al. (\cite{stark}) survey
and no additional intrinsic absorption. The photon index $\Gamma$ was
calculated individually for each object with the hardness ratios tabulated in
the RASS-BSC. The method is described in detail in Schartel et
al. (\cite{schartel}). 

$\Gamma$ and \NH\ were also determined simultaneously with
the two hardness ratios, which yields
large uncertainties for low fluxes. 
We compared the freely fitted \NH\ with the Galactic \NH\ and found that 
for all objects the Galactic \NH\ lies inside the $2\sigma$ errors of the
freely fitted \NH. 

A common feature of QSO and Seyfert 1 ROSAT spectra is 
the so-called  soft X-ray excess, an
additional thermal component of AGN spectra in the extreme soft X-ray and EUV
region (Schartel et al., \cite{schartel}). This feature is not characteristic
of BL\,Lac objects as our analysis and previous studies (Lamer et al.,
\cite{lamer}, Padovani \& Giommi, \cite{padovanigiommi}) have shown. 
The spectral
behaviour of BL\,Lac objects in the soft X-ray region is well described 
by a simple model and the determination of their flux is unambiguous. 
The
conversion factors of the presented objects for the hard ROSAT band differ by
only $15\%$ due to the varying Galactic absorption in the surveyed area. 
For 35 of 39 objects with Galactic $\NH < \rm 2.5\cdot
10^{20}\,cm^{-2}$ the difference is only $7\%$.

Luminosities are calculated under the assumption of
isotropic X-ray emission which need not be correct for BL\,Lac objects.

\subsection{Optical spectra and redshift determination \label{zbestimmung}}
 Most of the BL\,Lac objects discussed in this paper have been
 previously identified as such but without a reliable redshift.
We took new
high signal to noise spectra in order to determine the redshifts of these
objects and
of additional unpublished BL\,Lac candidates which fulfill the sample
criteria. In March 1997 the focal reducer MOSCA at the Calar Alto 3.5m
telescope was used for four nights to
take spectra with spectral resolutions of 6\,\AA\ ($\rm 4150 - 6650\AA$) and
12\,\AA\ ($\rm 4250 - 8400\AA$). The 2\,min
Johnson R aquisition frames were used for photometry and for checking the
BL\,Lac host galaxy. The B magnitude in Table~\ref{basistabel} is drawn from
this photometry for the newly discovered objects and has an estimated
uncertainty of ca. 0.4\,mag. B values for already catalogued objects are taken
from the literature (Nass et al., \cite{nass}).
In April 1997 we took additional spectra for two nights 
at the WHT on La Palma with the ISIS 
double spectrograph which yielded a spectral resolution of 10\AA\ between
3700\,\AA\ and 8900\AA.
More details can be found in the observation log 
(Table~\ref{observationlog}).
In the case of RX\,J1031.3+5053 we analyzed 
an additional spectrum taken with the old focal reducer at the 
3.5m telescope on Calar Alto. For B2\,1215+30 and PG1218+304 we got new
spectroscopically determined redshifts from Perlman et al., in preparation.
The redshift
of RX\,J1111.5+3452 was kindly provided by Schwope and Hasinger,
private communication. 

\begin{figure} %einspaltig
\epsfig{file=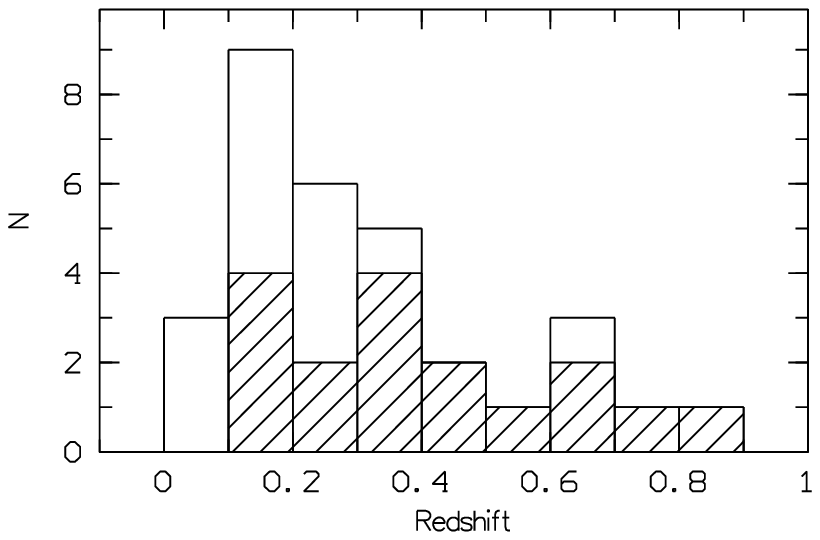,width=8.6cm}
\caption[]{\label{zhisto} Redshift distribution of the 31 BL\,Lac objects 
  from the complete sample. There are 4 more objects in the complete sample 
  without reliably determined redshift. Objects with $\alphaox < 0.91$ are
  shaded}
\end{figure}

The optical data were reduced within the MIDAS
package with an
optimized Horne algorithm for the extraction of the spectrum which also
provides  a $1\,\sigma$\ noise estimate of each spectrum based on Poisson
statistics. A careful
flat field correction was performed.
Nevertheless fringing patterns 
remained in the MOSCA spectra above 7000\,\AA. The ISIS spectra
did not have this problem but they suffered from a variable flat field
below 5000\,\AA. In summary both instruments did not provide optimal results 
but the observations supplemented each other.  
Since telescope time is scarce we did not 
in general reobserve objects with known 
redshifts, 
but in two cases where we did so
we found a 
redshift different from the 
published value.
Therefore it would be strongly 
desirable to reobserve at least these
BL\,Lac objects for which the redshift is marked 
in the literature as tentative or
ambiguous.

\begin{table}
\caption[]{Integration times and redshift confidence 
  for the BL\,Lac objects observed March 1997 with
  the Calar Alto 3.5m (CA3.5) telescope and April 1997 with the 4.2m telescope
  (WHT) on La Palma \label{observationlog}}
\begin{center}
\begin{tabular}{lrrc}
Name & CA3.5 & WHT & redshift$^{1}$ \\
     & $[min]$ & $[min]$\rule[-2mm]{0mm}{2mm} & \\
 \hline
RX\,J0809.8+5218\rule{10mm}{0mm}  & 60 & & C \\
RX\,J1008.1+4705  & 45 & & C \\
RX\,J1031.3+5053 &  55$^{2}$ &  & T \\  
RX\,J1058.6+5628 &  & 45 & C \\
RX\,J1248.3+5820  &  & 30 &  \\
RX\,J1302.9+5056  & 95 & 60 & C \\
RX\,J1324.0+5739 &  25 & & C \\
RX\,J1353.4+5601  & 30 & & T \\
RX\,J1404.8+6554  & 30 & & C \\
RX\,J1410.5+6100 &  40 & & C \\
RX\,J1422.6+5801 &  85 & 60 & C \\
RX\,J1436.9+5639 &  30 & 60 &  \\
RX\,J1451.4+6354 &  30 & & C \\
RX\,J1456.0+5048  & 70 & & C \\
RX\,J1458.4+4832 &  60 & & C \\
RX\,J1517.7+6525 &  30 & &  \\
1ES\,1533+535  &    60 & 60 & P \\
RX\,J0832.8+3300 &  105 & & C \\
B2 0906+31    &   30 & & C \\
RX\,J0915.8+2933 &  60 & &  \\
RX\,J1230.2+2518 &  & 60 &  \\
RX\,J1237.0+3020 &  120 & 60 & C \\
RX\,J1241.6+3440  & 45 & 60 &  \\
RX\,J1326.2+2933 &  30 & & C \\
RX\,J1631.4+4217  & 56 & & C \\
    
\end{tabular}
\end{center}
$^{1}$ redshift C: certain, T: tentative, P: possible\\
$^{2}$ Includes 25min  from March 1994 \\

\end{table}

The characterizing feature of optical BL\,Lac object spectra is a nonthermal
featureless continuum which is well described with a single power law. A
second component is contributed by the host galaxy.
A redshift determination is possible only with this component. 
The host galaxies are in the majority giant elliptical galaxies
(Wurtz et al., \cite{wurtz}). 
These galaxies have strong absorption features which are produced 
by the sum of their stellar content. Strong absorption lines in
the optical wavelength region of
elliptical galaxies and also of most spiral galaxies are an iron feature at 
3832\,\AA, Ca H and K (3934\,\AA\ and 3968\,\AA, respectively), the G Band at
4300\,\AA, Mg\,Ib at 5174\,\AA\ and the Na\,I\,D doublet (5891\,\AA). Another
important feature is the absorption edge at 4000\,\AA\ which has a contrast of
more than $40\%$ in galaxies with a late stellar population. 

The interstellar gas contributes weak narrow emission features to the
 spectrum. They are weak in normal elliptical galaxies but
the most powerful elliptical galaxies, cD galaxies, often show LINER
properties and therefore have stronger emission lines.

For a reliable redshift determination we adopted the rule that 
at least two spectral features must be unambiguously discernible.
The strength of the absorption features depends on the
flux ratio between the nonthermal featureless continuum and the host galaxy
contribution. The strongest absorption line in the optical wavelength region
of galaxies is the Ca K line with a typical equivalent width of $\rm
W_{\lambda} = 7\AA$ (Kennicutt, \cite{kennicutt}). 
In our high signal to noise spectra absorption lines can
be detected down to ca. $\rm W_{\lambda} = 0.3\AA$ (the spectra of the optical
faint objects have lower S/N and thus a higher detection limit for lines). 
Therefore it becomes impossible to detect Ca K in our spectra
if the host galaxy portion is lower than 0.05 or the
magnitude difference is larger than 3.5\,mag between host galaxy and
nonthermal continuum. 

For higher redshifts Ca H and K move out of the
optical wavelength region, while 
absorption lines produced by the interstellar gas of
the host galaxy become observable. 
The strength of these lines depends on the
relevant ion column density and is in principle independent of the stellar
content 
of the galaxy. It depends on the gas column density, its ionization state, 
metal
composition, and optical depth. These lines are the typical quasar absorption
lines. The strongest lines for objects around $\rm z = 1$ are the MgII doublet
(2796.4\AA\ and 2803.5\AA), MgI 2853.0\AA, the two FeII lines (2382.8\AA\ and 
2600.2\AA), and FeI 2484.0\AA, which have
equivalent widths up to several \AA\ (Verner et al., \cite{verner}). 
It is difficult to distinguish these
lines from ones produced by intervening material.
Therefore redshifts derived on this basis
are lower limits rather then firm values as
derived from  lines produced by the stellar population.
Moreover, 
since the gas content of elliptical galaxies is low
the expected  equivalent widths are small
and are possibly at the detection limit of our spectra.

\begin{figure} %einspaltig
\epsfig{file=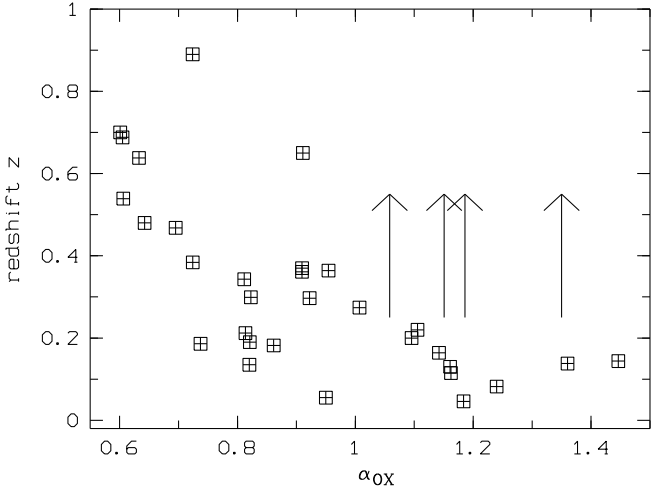,width=8.6cm}
\caption[]{\label{zalphaox} Redshift versus \alphaox. 
The objects with unknown
  redshifts are indicated with arrows}
\end{figure}

The ratio between AGN continuum and host galaxy contribution can also be
constrained by analyzing the spatial profile of the direct image of the
BL\,Lac object. We have compared the BL\,Lac images from the 2\,min R
aquisition frames with star images on the same frame.
We found significant deviations from
a stellar appearance for all BL\,Lac objects for which we could determine the
redshift including the objects with higher redshift $\rm z > 0.5$. Most of the
objects without reliable redshift revealed no extension and
could be well described with a stellar appearance. 
Three of these objects are optically bright, $\rm B < 16.5$. A bright host
galaxy with $M_{B} = -21$ at $\rm z = 0.25$ 
would appear already 3.5\,mag fainter than the nucleus. Therefore we assume 
$\rm z \ge 0.25$ for these optically brighter objects. 
Fainter host galaxies would
decrease the lower limit for the redshift. The constraints for the optically
fainter objects are stronger. We expect $\rm z \ge 0.6$, when the 4000\,\AA\
break moves out of the R band.

In Fig. \ref{spektren} the optical spectra are shown. Many spectra were taken
under non-photometric conditions. Therefore the spectrophotometric use of the
spectra is strongly limited. For some objects we show two spectra taken with
different telescopes. In general, they do not join due to different
atmospheric conditions and intrinsic variability. The spectra are smoothed
with a Gaussian filter of width  appropriate to the FWHM 
of the instrument used. The most important
features for the determination of the redshift are marked. We present only
objects for which we could determine a new redshift. The other spectra contain
no real information and show only a high signal to noise power law with some
telluric absorptions. 
Some of the spectra show severe deviations from a power law, they are produced
by a large contribution from the host galaxy. In other spectra features used
for the redshift determination have the same strength as telluric and residual
flat field features. In such cases the $1\,\sigma$ noise estimate is
helpful. The certainty of spectral features significantly increases if they
are found in spectra taken with different grisms and telescopes. Comments on
individual objects are given in the appendix. They are summarized in
Tab. \ref{observationlog}.

\begin{figure} %einspaltig
\epsfig{file=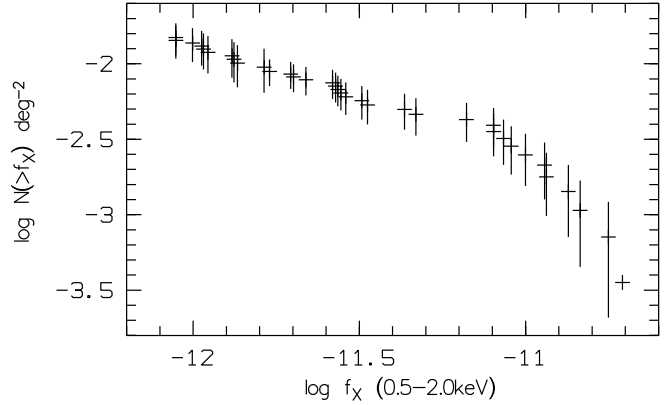,width=8.6cm}
\caption[]{\label{lognlogs} $\log N(> S) - \log S$ distribution for the 
  complete
  sample. Only objects above the completeness level have been taken into
  account }
\end{figure}

\section{Analysis \label{analysis}}
\subsection{Comparison to existing samples \label{zcompare}}
The $\log N(> S) - \log S$ distribution (Fig. \ref{lognlogs}) is independent 
of the subtle redshift determination process for BL\,Lac objects. Our sample
shows a steep slope consistent with the Euclidean value of $-1.5$ down to
$\rm f_{X} = 8\cdot 10^{-12}\ergscms$. 
Below this flux we have an extremely flat
slope of $-0.5$. The flattening cannot be due to incompleteness since the HRX
is fully identified down to the flux limit. 
If  the statistical uncertainties and the
different energy ranges\footnote{assuming a power law with $\Gamma = -2.1$ 
  the abscissa in Fig. \ref{lognlogs}
  values have to be shifted by 0.36 to be comparable with the full ROSAT
  energy range ($0.1 - 2.4$\,keV) or 0.20 for the {\em Einstein} range ($0.3 -
  3.5$\,keV), respectively} 
are taken into account the 
distribution is compatible with 
the surface density we previously published
(Nass et al., \cite{nass}) which was
derived from a larger sky area. However there are
deviations to the EMSS sample of
BL\,Lac objects (Maccacaro et al., \cite{maccacaro}). The
biggest difference arises 
around our turnover point near
$\rm f_{X} = 8\cdot 10^{-12}\ergscms$.
This difference could be caused by the fact that the EMSS
discarded the targets of
the pointed observations which were used to select the serendipitous X-ray
sources. Some targets were X-ray bright BL\,Lac objects. 
There are further differences between the EMSS and the RASS sample, e.g. the
spectral window used for the selection process. In principle, theses
differences could be responsible for the small distinctions of  $\log N(> S) -
\log S$ in both surveys.
Near our flux limit
the space densities from the two independent surveys are fully consistent.
The EMSS found a
turnover point in the $\log N(> S) - \log S$ curve around 
$\rm 10^{-12}\ergscms$
(Maccacaro et al., \cite{maccacaro}).
This cannot be verified with our sample since it is near
our flux limit.

\begin{figure*} %zweispaltig
\epsfig{file=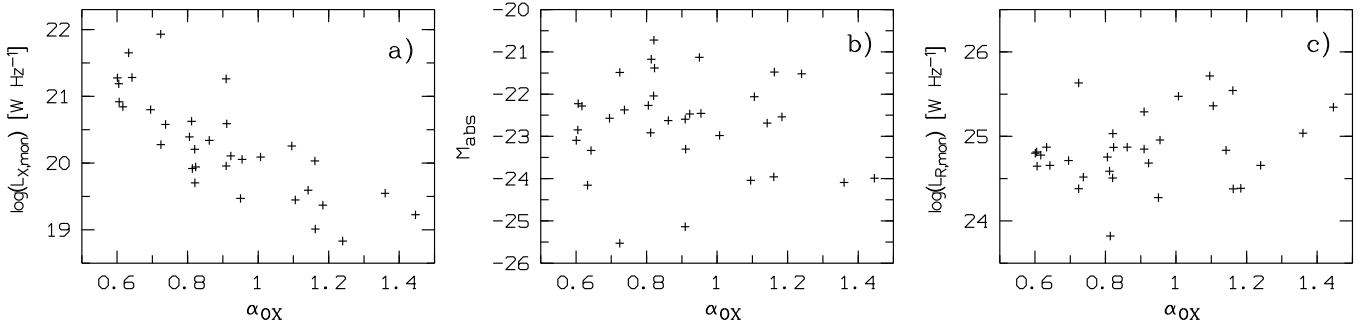,width=18.0cm}
\caption[]{\label{lumis} Luminosities in three wavelength bands versus
  \alphaox. a) monochromatic X-ray luminosity at 2keV b) absolute optical
  magnitude c) monochromatic radio luminosity}
\end{figure*}

Our BL\,Lac sample is selected from a sky area 
approximately three times
larger than the effective sky area of the EMSS. This 
presumably explains the  
fact that we found three 
objects with $\rm z < 0.1$ and
the EMSS none. The selection process of the EMSS 
might also have played a role for
bright nearby objects. But it is surprising that we found more objects
with higher redshifts $\rm z > 0.4$ and also objects with higher redshifts
than present in the EMSS ($\rm z = 0.638$). The redshift distribution is
also flatter 
% (with a median of $\rm z = 0.33 ??$) 
but there is still a
bump around $\rm z = 0.2$. This bump in the distribution nearly disappears if
only the extremely X-ray dominated BL\,Lac objects with $\alphaox < 0.91$ 
(the median of
\alphaox) are taken into account. 
The low redshift objects are excluded with this criterion 
and the
distribution is more similar to the radio selected BL\,Lac objects from the
1\,Jy sample (Stickel et al., \cite{stickel}). It is 
improbable that the 
objects with unknown
redshifts,
if included,   %NGD%  ??? 
would destroy this distinction. Most objects with unknown
redshift have $\alpha_{OX} > 0.91$ and we have already argued in
Sect. \ref{zbestimmung} 
for medium redshifts $\rm 0.25 < z < 0.7$ for these objects. The
bump around $\rm z = 0.2$ will remain and the objects with unknown redshift
cannot be low redshift
objects with low \alphaox. We will come back to this important result after
 the calculation of the luminosity function.

Although the redshift distribution of the extremely X-ray dominated objects
is rather flat, somewhat similar to the distribution of the 
1\,Jy BL\,Lac sample (Stickel et al., \cite{stickel}), 
the analogy cannot be driven too far. 
The majority of the higher redshifts $\rm z > 0.4$ in
the 1\,Jy sample were determined with emission lines and the strongest
lines often had equivalent widths around the 5\AA\ limit. In our RASS selected
sample all redshifts are determined with absorption features and the
equivalent widths of the few
emission lines which have been detected in some objects (see Appendix) are
far below 5\,\AA. Furthermore some of the emission
lines in the 1\,Jy sample are resolved and broad ($>500\kms$) which is a
characteristic feature of Seyfert 1 or QSO activity. All emission lines in
our objects are unresolved ($<500\kms$) and fully consistent with 
an origin in HII-regions in the host galaxy
--- no Seyfert nucleus is necessary to account 
for these emission lines. The absence of emission lines seems to
be a common feature of the optical spectra of X-ray selected BL\,Lac objects
since the spectra of the EMSS objects 
also show no 
emission lines (Morris
et al., \cite{morris}). This distinction between X-ray and radio selected 
BL\,Lac objects can be interpreted as an
indication of differing parent populations in X-ray and radio selected
BL\,Lac objects.

\begin{table*}
\caption[]{Cosmological evolution parameters and $\langle V/V_{max}\rangle$
for the complete sample \label{evolspec}}
\begin{center}
\begin{tabular}{lrrcrcc}
 & N & $\beta$ & $\beta$, $2\sigma$\ range & $\gamma$ & 
$\gamma$, $2\sigma$\ range & $\langle V/V_{max}\rangle$ \\
\hline
unknown redshifts omitted & 31 & $-4.0$ & $-7.6 < \beta < +0.2$ & $-4.9$ & 
$-13.6 < \gamma < +0.2$ & $0.40\pm 0.06$\\
without 1ES\,1533+535 & 30 & $-4.0$ &  $-8.1 < \beta < +0.7$ & $-4.6$ & 
$-13.8 < \gamma < +0.5$ & $0.41\pm 0.06$\\
unknown redshifts set to $\rm z = 0.3$ & 35 & $-3.7$ & $-7.2 < \beta < +0.3$ &
$-4.3$ & $-12.0 < \gamma < +0.2$ & $0.41 \pm 0.05$\\
unknown redshifts set to $\rm z = 0.3$, $\alphaox > 0.91$ & 18 &  & &
 &  & $0.48 \pm 0.08$\\
unknown redshifts set to $\rm z = 0.3$, $\alphaox < 0.91$ & 17 &  & &
 &  & $0.34 \pm 0.06$\\
unknown redshifts set to $\rm z = 0.7$ & 35 & $-3.2$ & $-6.4 < \beta < +0.6$ &
$-3.7$ & $-10.9 < \gamma < +0.5$ & $0.42 \pm 0.05$\\
unknown redshifts set to $\rm z = 0.7$, $\alphaox > 0.91$ & 18 &  & &
 &  & $0.49 \pm 0.08$\\
unknown redshifts set to $\rm z = 0.7$, $\alphaox < 0.91$ & 17 &  & &
 &  & $0.34 \pm 0.06$\\

\end{tabular}
\end{center}

\end{table*}

\subsection{Luminosity function}
Redshifts are available for 31 ($89\%$) of the 35 BL\,Lac objects forming 
a flux limited sample. This high portion of objects with known redshifts
allows the determination of a luminosity function. Possible biases and
selection effects introduced by the objects with unknown 
redshifts, 
which have been excluded from the 
luminosity function, are discussed later.

Determination of a luminosity function means 
taking all objects above the
flux limit, 
counting the objects above a given luminosity, and 
dividing the
number by the volume surveyed for these objects. The BL\,Lac sample is
distributed on two subareas with different flux limits which do not overlap
(see Sect. \ref{sampledefinition}).
The results of each area are  combined with the methods given in 
Avni \& Bahcall (\cite{avnibahcall}).

A first step to evaluate the contribution of each object to the
luminosity function is the calculation of the maximal redshift $\rm z_{max}$ 
where this
object would be detectable with the given flux limit. For this purpose a
K-correction is assumed which depends on the individually determined power law
spectra. It should be noted that it is not clear that the power law can be
extrapolated as far as it would be needed by the largest $\rm z_{max}$ 
which are around $\rm z = 2$.

Finally a luminosity function is obtained with this procedure which is
determined 
between the minimal and maximal redshift of the sample, 
but which assumes no
evolution. However, the objects are not evenly distributed in the volume, 
 $\langle V/V_{max}\rangle$ being $0.40\pm 0.06$ rather than $0.5$ as
expected for a uniform distribution.
It is common practice to determine luminosity functions under the assumption
of $\langle V/V_{max}\rangle = 0.5$ and a redshift dependent cosmological 
evolution. We have investigated 
density evolution with the parameterization $\rho(\rm{z}) = \rho(0)\cdot
(1+\rm{z})^{\beta}$ and luminosity evolution in the form $L_{X}(\rm{z}) =
L_{X}(0)\cdot (1+\rm{z})^{\gamma}$. $\langle
V/V_{max}\rangle = 0.5$ is achieved for $\beta = -4.0$ with $\pm 2\sigma$
limits of $-7.6$ and $+0.2$ and for luminosity evolution we obtain 
$\gamma = -4.9$
with $\pm 2\sigma$ limits of $-13.6$ and $0.2$
(see also Table~\ref{evolspec}).
 If 1ES\,1533+535 is omitted
from the analysis, the object with the highest but also most uncertain
redshift, the evolution parameters $\alpha$\ and $\gamma$\ are slightly shifted
to stronger negative evolution.
We now consider 
how  this result 
might be affected by the objects with 
unknown redshift. In Sect. \ref{zbestimmung} we have already argued that these
objects  have a high probability of high
redshift. For a check we have adopted $\rm z = 0.3$ and $\rm z = 0.7$ for
these objects and repeated the analysis. The changes for the evolution
parameters are
small. In both cases the $\pm 2\sigma$ limits are decreased
because the statistics are better, 
but otherwise there is little change.
These results can be generalized, if some of our tentative redshifts are not
correct. In such a case the evolution parameters will only slightly be
altered. However, the form of the luminosity function could be changed. In
particular, if our sample consists several objects with $\rm z > 1$ the
luminosity function will be enlarged to higher luminosities.

\begin{figure} %einspaltig
\epsfig{file=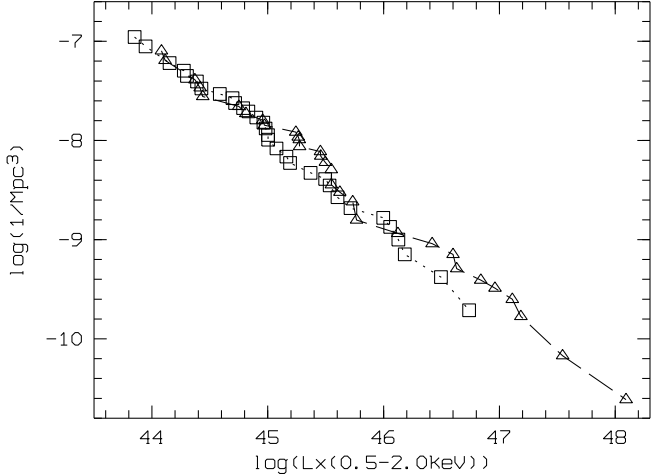,width=8.6cm}
\caption[]{\label{cumlfunc} Cumulative luminosity function for RASS selected
  BL\,Lac objects evolved to $\rm z = 0$ with the assumption of
  density evolution (marked with squares) and luminosity evolution
  (triangles).
  Objects with unknown redshifts are not considered}
\end{figure}

\begin{figure} %einspaltig
\epsfig{file=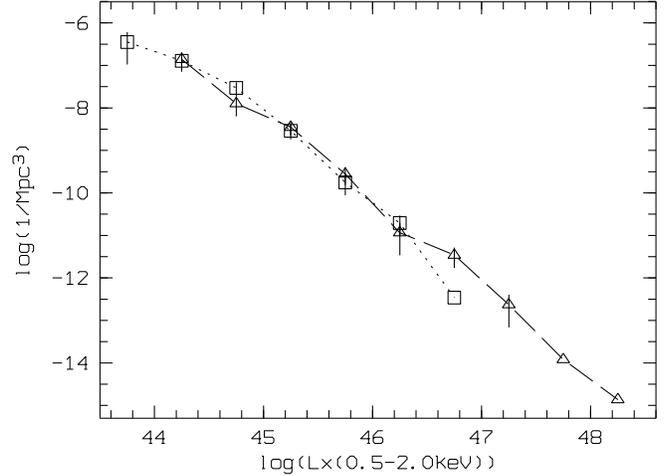,width=8.6cm}
\caption[]{\label{diflfunc} Differential luminosity function for RASS selected
  BL\,Lac objects evolved to $\rm z = 0$, same coding as in
  Fig. \ref{cumlfunc}. The bin size of the x-axis is 0.5, the y-axis 
gives the space density per luminosity interval of $\rm 10^{44}\,ergs\,s^{-1}$}
\end{figure}

In principle, our analysis of the luminosity function of RASS selected BL\,Lac
objects has given similar results compared to the analyses performed with
EMSS selected objects (Morris et al., \cite{morris} and Wolter et al.,
\cite{wolter}). Our sample shows negative evolution, too, although in a milder
form. The complete RASS sample contains 13 more objects than the sample 
presented in
Morris et al. (\cite{morris}) and 3 more than in Wolter et
al. (\cite{wolter}). Therefore the statistics are a little better,
nevertheless negative evolution is not ruled out on the $2\sigma$ level if we
adopt $\rm z = 0.3$ or $\rm z = 0.7$ for the objects with unknown 
redshift. Because most
of these objects are optically bright, redshifts closer to $\rm z = 0.3$
are probable.
A detailed comparison of the cumulative luminosity function, shown in
Fig. \ref{cumlfunc},  with previously published
functions is limited by the different evolution parameters which alters the
space densities. We cannot find strong deviations, the most important
difference is the extension of the RASS luminosity function to higher
luminosities by at least one order of magnitude. 
The differential luminosity function, shown in Fig. \ref{diflfunc}, can be well
described 
by a single power law with no significant bump and 
with a flat index $-1.9$. 
The statistics are too low for building redshift
bins,
so this cannot be used to distinguish between density
and luminosity evolution. Because luminosity evolution foresees extremely
powerful BL\,Lac objects at $\rm z = 0$ which are not observed,
we favour density evolution.

\subsection{Evolutionary behaviour and spectral energy distribution}
The $V/V_{max}$ variable, first introduced by Schmidt (\cite{mschmidt}), can
itself be used to discuss the cosmological evolution of objects. $V/V_{max}$
depends mainly on the ratio between observed flux and the survey flux
limit,
with a weaker dependence on the observed redshift
and cosmological model.

Motivated by the contradicting results about the evolutionary behaviour in
X-ray and radio selected BL\,Lac objects we divided our sample into two
subsamples according to the X-ray dominance in the spectral energy
distribution. The extremely X-ray dominated objects with 
$\alphaox  < 0.91$
have $\langle V/V_{max}\rangle = 0.34\pm 0.06$ and the
intermediate objects have $\langle V/V_{max}\rangle = 0.48\pm 0.08$. 
With a Kolmogorov-Smirnov test we obtain a probability of 0.19 for the null
hypothesis that both subgroups are taken from the same parent
sample. 
The $\langle
V/V_{max}\rangle$ statistic develops with $1/\sqrt{12N}$.
Again the effect of objects with unknown redshift must be checked. 
They are unevenly distributed in this
case (see Fig. \ref{zalphaox}). 
All these objects have $\alphaox > 0.91$ with one object having exactly the 
median value. The changes introduced by the 4 objects are small if we adopt
$\rm z = 0.3$ or $\rm z = 0.7$ for them (see Table~\ref{evolspec}). 
Statistically the difference of $\langle V/V_{max}\rangle$ is only significant
at the $2\sigma$ level. This result matches well with previously published
analyses, leading us to believe in the reality of this observation. 
Apparently contradictory results relating to cosmological evolution
in X-ray selected (Morris et al., \cite{morris}) 
and radio selected (Stickel
et al., \cite{stickel})  %NGD%  word missing ???
BL\,Lacs are 
resolved in our sample.
The bump around $\rm f_{X} = 8\cdot 10^{-12}\,ergs\,cm^{-2}\,s^{-1}$ in the
 $\rm \log N(> S) - \log S$ distribution (Fig. \ref{lognlogs}) is also
mainly produced by the objects with low \alphaox. In a  $\log N(> S) - \log S$
plot with only these objects the bump is exaggerated whereas for objects with
$\alphaox > 0.91$ it is insignificant (but does not totally disappear). 

In Sect. \ref{zcompare} we found a
distinction in the redshift distribution for the two BL\,Lac object subgroups
in Sect. \ref{zcompare}. Here the statistical significance is higher, a
Kolmogorov-Smirnov test yields a probability of 0.009 for the null
hypothesis. Even if $\rm z = 0.7$ is adopted for the unknown redshifts the
distinction between the two subgroups remains. All objects with $\rm z < 0.1$
have $\alphaox > 0.91$. Figure \ref{zalphaox} implies an anticorrelation
between \alphaox\ and z, but the 
significance of this relation
is reduced by the unknown redshifts. 
As for luminosities, both the uneven
redshift and the $V/V_{max}$ distribution 
have similar effects,
so that it can be stated that the
X-ray luminosity distributions  of 
X-ray dominated and
intermediate objects 
are clearly distinct, with a probability of 0.002 for a
common parent population.

In order
to look for additional confidence in the redshift distribution distinction
we verified other X-ray selected BL\,Lac
samples. If the EMSS BL\,Lac sample is divided
into equal halves according to the ratio between X-ray and radio flux, the
more X-ray dominated subgroup has higher redshifts, too. The {\em Einstein}
Slew Survey sample (Perlman et al., \cite{eslew}) is not well suited for such a
comparison since the redshift 
is missing for many of the X-ray dominated objects. But again, no $\rm z <
0.1$ object  is found among the objects with $\alphaox < 0.91$, although there
are several such objects among the intermediate objects.

\subsection{Selection effects}
A selection effect described by
Browne \& March\~{a} (\cite{browne}) 
is that low
luminosity and nearby BL\,Lac objects are difficult to distinguish from their
host galaxies because their nucleus is dominated by starlight and does not
fulfill the spectroscopic requirements for a featureless optical spectrum. 
This selection bias cannot be responsible for the higher redshifts in objects
with low \alphaox. Bright objects with small redshifts are also missing in
this subgroup. And there is no reason why we should have misidentified all low
redshift objects with low \alphaox, but did select such objects with high
\alphaox.

In  X-ray surveys faint BL\,Lac objects can also be lost in the extended
X-ray emission of a rich hosting galaxy cluster. This is only possible if the
BL Lac object is considerably fainter in X-rays compared to the surrounding
cluster. Therefore only the X-ray faint BL Lac objects can be overpowered by
clusters since X-ray clusters reach only X-ray luminosities of several
$\rm 10^{45}\,ergs\,s^{-1}$ (Schindler et al., \cite{schindler}).
Therefore consequences of this selection
effect are comparable to the problems caused by strong host galaxies. Again,
the distinction between objects with low and high \alphaox\ 
is not readily explicable by means of this effect.

\section{Discussion \label{models}}
We have determined the luminosity function of RASS
selected BL\,Lac objects which has a larger luminosity range than previously
published luminosity functions.  The analysis of the evolutionary behaviour
confirmed earlier statements about a negative evolution of X-ray selected
BL\,Lac objects. In our sample the negative
cosmological evolutionary  behaviour is restricted to the extremely X-ray
dominated objects 
with $\alphaox < 0.91$. Further hints for two distinct populations can be
found in the $\rm \log N(> S) - \log S$ distribution. The bump around $\rm
f_{X} = 8\cdot 10^{-12}\,ergs\,cm^{-2}\,s^{-1}$ is mainly caused by the
extremely X-ray dominated objects.
In our sample the 
 BL\,Lac objects with low \alphaox\ are a high X-ray luminosity
population with 
few low luminosity objects. 
There is no clear distinction 
in optical and radio luminosity 
between the two
subgroups. This 
is 
not compatible with the assumption that X-ray
selected BL\,Lac objects are the low luminosity part of a single BL\,Lac
population (Urry \& Padovani, \cite{urrypadovani}).
Still, the radio luminosity of most RBLs of the
1\,Jy sample exceeds the luminosities found in our sample.

The characteristic features of BL\,Lac objects are often explained with a
relativistic jet aligned to our direction.
While this view is widely accepted,
more detailed specifications  about the physical nature of the jet and the
parent population, the objects with jets not pointed towards our direction, are
still in debate. Here we want to mention some consequences of our new
results on existing models, in particular the different redshift distributions
of objects with low and high \alphaox.

\subsection{\label{parentpopulation} Unified schemes and the parent population}
The unified scheme for BL\,Lac objects tries to identify X-ray and radio
selected BL\,Lac objects with intrinsically
 identical parent populations in
which only some physical parameters of the jet vary (e.g. beaming properties,
angle $\Theta_{C}$ between line of sight and jet). If only one parent
population is responsible for BL\,Lac objects the luminosity function of
BL\,Lac objects has to be consistent with the luminosity function of the
parent population under the assumption of physically reasonable beaming
properties. Padovani \& Urry (\cite{padovaniurry}) investigated whether
Fanaroff-Riley type I (FR\,I) radio galaxies can be a plausible parent
population. Assuming a simple beaming model for jets in FR\,I galaxies
described in Urry \& Shafer (\cite{urryshafer}) they calculated the
luminosity function of X-ray selected BL\,Lac objects. In principle they
predicted a beamed luminosity function that has a double power law with the
same power law index as the parent population above a certain break point and
a considerably flatter slope below this point. The predicted break point lies
at ca. $\rm L_{X} = 10^{43}\,ergs\,s^{-1}$ and is outside the covered range 
of our luminosity function. 

The power law index ($-1.9$) of our differential luminosity function is 
only slightly flatter  than the
index of FR\,I galaxies as determined by Padovani \& Urry
(\cite{padovaniurry}) ($-2.1$). 
Given the uncertainty in the luminosity function of the BL\,Lac objects
and the (similar)  uncertainty in that of the  FR\,I galaxies,
this difference is not significant.
Since negative evolution of BL\,Lac
objects was not in popular discussion, Padovani \& Urry (\cite{padovaniurry})
discuss only the effects of positive evolution. To adjust their model for the
extremely high X-ray luminosities found in our sample (roughly one order of
magnitude higher than in the calculation of their paper) stronger beaming must
be assumed with bulk Lorentz factors $\gamma > 7$ which is comparable to
values found in radio selected BL\,Lac objects. 

The assumption of one parent population for BL\,Lac objects is not commonly
accepted. Wall \& Jackson (\cite{walljackson}) allowed radio galaxies of
Fanaroff-Riley type I and II to generate BL\,Lac objects if their jets are
aligned towards us. With this scheme they could reproduce 5\,GHz source
counts of different extragalactic radio source classes, but they had to adopt
higher Lorentz factors than determined by Padovani \& Urry
(\cite{padovaniurry}). The method of Wall \& Jackson (\cite{walljackson})
cannot be easily transformed to our dataset because the radio data of our
X-ray selected sample are inhomogeneous. More detailed radio observations are
necessary to investigate whether the differing sample properties of BL\,Lac
objects below and above $\alpha_{OX} = 0.91$ can be attributed to different
parent populations.

\subsection{Jet models and selection effects}
Studies of BL\,Lac objects have shown that X-ray surveys
are an efficient tool to find BL\,Lac objects, and that X-ray selected BL\,Lac
objects outnumber radio selected objects. This is 
just what would be expected if
X-ray selected objects are intermediate in orientation between radio selected
BL\,Lac objects and radio galaxies. A physically justified explanation for
this is an accelerating jet model where the X-rays arise from regions closer
to the core than the regions emitting the radio emission. Applying this
``wide jet'' model,
 Ghisellini et al., \cite{ghisellini}, estimate that radio selected
BL\,Lac objects have an opening angle $\Theta < 15\degr$ (for which the
observer would see comparable radio emission), 
that  X-ray selected objects have
$15\degr < \Theta < 30\degr$ and 
that objects with $\Theta > 30\degr$ appear 
to the observer as
normal radio galaxies.
Celotti et al., \cite{celotti}, used
this picture to calculate the luminosity functions of X-ray and radio selected
BL\,Lac objects and in principle they were able to reproduce the observed data.
However, this model has problems 
explaining our result of a
redshift distribution 
dependence on the multifrequency spectrum. The most common
objects are the intermediate objects, the extreme X-ray and radio dominated 
objects have lower space density. Perhaps the model can be adjusted to the
correlation between \alphaox\ and redshift, but then additional free parameters
are needed.

As an alternative to the ``wide jet'' model Giommi \& Padovani
(\cite{giommipadovani}) have suggested that the distinction between X-ray and
radio selected BL\,Lac objects is caused by the location of the high energy 
cutoff of the synchrotron emission 
% (their model does not need the assumption of synchrotron emission), 
and that the two populations are fundamentally the
same. In this scenario, X-ray selected BL\,Lac objects are intrinsically
rarer, contrary to the standard view. The larger X-ray numbers originate
solely from selection effects. 
With this scheme the authors can reproduce the
redshift and luminosity distribution in the X-ray and radio band for X-ray and
radio selected BL\,Lac samples. 
However, this interesting scheme does not explain why low redshift objects are
not found among the extremely X-ray dominated objects. In our flux limited
sample the intermediate objects have the lowest X-ray luminosity. 
Regarding the optical and radio luminosity there is no clear division into two
subgroups. 

\subsection{Variability in the spectral energy distribution}
Multiwavelength observations of outbursts of BL\,Lac objects could be helpful
for understanding the anticorrelation between redshift and
\alphaox. Coordinated observations from the optical to the extreme
$\gamma$-region (TeV-region) of the X-ray bright BL\,Lac objects Mrk\,421
(Macomb et al., \cite{macomb}) and Mrk\,501 (Catanese et al., \cite{catanese})
revealed strong correlated flux variations in the X-ray and TeV region whereas
the correlated variability in the optical region was only small.
In March 1997 Mrk\,501 varied by a factor of 4 in flux 
between $2-10$\,keV and less than
$20\%$ in the U-band. During the outburst of Mrk\,501 the peak of $\nu
f_{\nu}$ in the X-rays, which is interpreted as the high synchrotron energy cutoff, was located at energies of $\sim 100$\,keV. Catanese et
al., \cite{catanese} conclude from these observational results that the
maximum electron Lorentz factor of Mrk\,501 is of the order $\rm \gamma_{max}
\sim 10^{6}$. Macomb et al., \cite{macomb}, propose that a similar outburst in
Mrk\,421 could be caused by a change in the upper energy cutoff in the
relativistic electron distribution without affecting either the magnetic field
or the normalization of the electron energy distribution.

The interesting aspect of these observations is that Mrk\,501 and Mrk\,421 are
objects with $\alphaox$ typical for intermediate objects,
while in times of high
activity their \alphaox\ decreases to values characteristic for the extremely
X-ray dominated objects in our sample. Buckley et al., \cite{buckley}, suggest
that extremely X-ray dominated BL\,Lac objects also have high synchrotron
energy cutoffs. Furthermore we suggest that the extremely X-ray dominated
objects with $\alphaox < 0.91$ are observed in a state of enhanced activity
whereas the intermediate BL\,Lac objects have lower energy cutoffs and they
would represent the quiescent population. Because the X-ray luminosities and
redshifts of the objects are higher in the 1\,Jy survey than for intermediate
BL\,Lac objects this speculative image is not consistent with the idea of a 
single BL\,Lac population. However, it offers a simple explanation for the 
anticorrelation between the redshift and X-ray luminosity and \alphaox. 

Kollgaard et al., (\cite{kollbeaming}) measured  
with radio methods the bulk Lorentz
factor $\gamma$ for several BL\,Lac objects.
They conclude that X-ray selected BL\,Lac objects have smaller 
values of $\gamma$ than radio selected BL\,Lac objects which is in
contradiction to our results in Sect.  \ref{parentpopulation} and of
Catanese et al., \cite{catanese}.  We note that we would classify almost
all X-ray selected objects used for these radio measurements as
intermediate, which would solve this discrepancy. 

\section{Summary and Conclusions}
We have shown that the unified models do not predict
the distinction between the two subgroups in our sample and 
additional free parameters would be necessary to adjust the model with the 
observations. 
Currently the observational constraints are of low statistical
significance. Two populations of BL\,Lac objects with
different physical origin are also consistent with the results. 
The dividing point,
$\alphaox = 0.91$, 
is the median of the sample and has no special physical
significance.

Our study has confirmed the negative evolution of X-ray selected BL\,Lac
objects and we find that this behaviour could be restricted to the
extremely X-ray dominated BL\,Lac objects. The cosmological evolutionary
behaviour of the intermediate objects is consistent with no evolution.
Since these two subgroups show a different redshift distribution the
interesting evolution time of X-ray dominated BL\,Lac objects is shifted
outside to $\rm 1.0 < z < 2.0$ 
(the $\rm z_{max}$ from the $V/V_{max}$ analysis). 
There is still no answer for this odd behaviour, which up to now has  
not been found in other AGN classes. Possibly the jets in these
objects, which are especially powerful in the X-rays, need a minimal time for
development. This speculation has gained some plausibility since the analysis
of our sample has shown that the relevant times lie farther away. Nevertheless
X-ray dominated BL\,Lac objects must have been considerably rarer between 
$\rm 2 < z < 3$, the time when QSO activity was at its highest point.
The distinct properties in the two subgroups are already revealed in the
bright part of the $\log N(> S) - \log S$ distribution. It is therefore
implausible that selection effects, as described by Browne \& March\~{a}
(\cite{browne}), are responsible for this distinction.
Nevertheless, BL\,Lac objects are ``susceptible'' to selection effects and
therefore, in future, galaxy clusters should be carefully analyzed in the
selection process.

The observations suggest the intermediate BL\,Lac objects as the
basic population. They have the lowest luminosity in X-rays and radio
wavelengths, and they have the highest space density. Both the extremely
radio and X-ray dominated BL\,Lac objects have higher luminosities. 
This led us to speculate about a beaming scenario. 
The spectral energy distribution of extremely X-ray dominated BL\,Lac objects
can be interpreted with a high energy cutoff of the synchrotron spectrum.
The high X-ray luminosity of objects with $\alphaox < 0.91$ can be explained
with large bulk Lorentz factor of relativistic electrons in the jets of these
objects. This result is compatible with conclusions from multiwavelength
observations of outbursts in Mrk\,421 and Mrk\,501. Therefore we suggest
that extreme X-ray dominated objects are observed in a state of enhanced
activity which would explain the anticorrelation between X-ray luminosity and
\alphaox.

Our investigation has shown the importance of the redshift parameter for the
study of samples of BL\,Lac objects. 
Advances in observing techniques have
made easier the determination of redshifts in optical spectra which
are devoid of strong features. 
In order to 
have stronger observational constraints it is necessary
to enlarge the sample
and to determine redshifts for
objects for which
this has not yet been possible. 
The uncertainty of the $V/V_{max}$
statistics depends on $1/\sqrt{12N}$. In the last years several projects
began to select BL\,Lac objects with combined X-ray - radio methods
(e.g. Wolter et al., \cite{pilot}). They will certainly yield valuable
results about intermediate objects and low flux BL\,Lac objects below the
adopted flux limit in this paper.
However, the extremely X-ray dominated BL\,Lac objects are rare and their
negative evolution leads to a very flat $\log N - \log S$. 
Their distinct population properties are an important tool to understand the
BL\,Lac phenomenon, 
and they can
be selected most preferably with the 
full sky coverage of the  RASS.

\acknowledgements{The ROSAT project is supported by the Ministerium f\"ur
  Bildung, Wissenschaft, Forschung und Technologie (BMBF/DARA) and by the 
  Max-Planck-Gesellschaft (MPG). We thank K. Molthagen for assistance 
  during observations.
 L. Wisotzki
  is thanked for providing his implementation of the Horne algorithm. 
  NB acknowledges support by 
  the BMBF under DARA 50\,0R\,96016.}

\appendix
\section*{Appendix}

The appendix gives comments on the
classification of some objects 
as  BL\,Lacs 
if the objects are optically extremely faint or
other peculiarities exist. For objects with subtle
optical spectral features details of 
the redshift determination are provided 
and tentative or ambiguous redshifts are marked.

\begin{description}
\item[RX\,J0809.8+5218] This object has an extremely strong interstellar
  Na\,I\,D absorption line, stronger than the absorption features from the
  host galaxy. There exists a second optical spectrum with higher resolution
  which confirms 
 our redshift. This object is also a member of the {\em
  Einstein} Slew Survey.
\item[RX\,J1031.3+5053] In March 1997 a 30min exposure with 6\AA\ spectral 
 resolution was taken. Two weak lines ($\rm W_{\lambda} = 0.4\AA$)
 were found 
 with the correct separation for Ca H and K.
 There is a suggestion of the G band in this spectrum
 but it is barely discernible from the noise.
 The Ca H and K lines are
 confirmed in a spectrum with a lower resolution (15\AA) spectrum from March
 1994, in addition a weak 4000\AA\ break is apparent. This object is also a
 member of the {\em Einstein} Slew Survey. We did not find the absorption
 features  indicated by Perlman et al. (\cite{eslew}) in our spectrum which led
 to a different redshift. This contradiction and the weakness of the absorption
 feature let us consider the redshift of $\rm z = 0.361$ as tentative although
 two spectra exist which supplement each other with our redshift.
\item[RX\,J1058.6+5628] In the red part of the WHT spectrum Na\,I\,D
  absorption is found at 6740\AA. It appears broad in accordance with its
  doublet character. In addition weak emission lines from [NII]6548, H$\alpha$
  and [NII]6584 can be found for this redshift. The strongest emission line,
  [NII]6584, has an equivalent width of $\rm W_{\lambda} = 0.9\AA$, 
  well below the
  classification limit. Ca H and K and the G band are strongly affected by
  varying flatfield corrections in the blue part of our WHT spectra, but they
  can be found at the appropriate place in spectra taken by Laurent-Muehleisen
  (\cite{laurent}). We consider this redshift therefore as secure. We cannot
  confirm the redshift given in March\~{a} et al. (\cite{marcha}) which 
  is based
  on [OIII] emission lines at 7000\AA\ where neither our spectrum nor spectra
  taken by Laurent-Muehleisen (\cite{laurent}) show emission lines.
\item[RX\,J1302.9+5056] This optically faint object is not visible on the
  digitized POSS I data and the HQS direct plates. The position is taken from
  the digitized POSS II data. The spectral features (MgII\,2798, Ca H and K
  and G band) used for the redshift
  determination have been found on  spectra taken with different telescopes.
\item[RX\,J1324.0+5739] This low luminosity object has very strong absorption
  features. The contrast of the 4000\AA\ break 
  is, at $24\%$, at the border of
  the classification limit. The available radio and X-ray properties are
  consistent with a BL\,Lac classification and we adopted it for our
  analysis. We note that spectra taken by  Laurent-Muehleisen
  (\cite{laurent}) have a stronger 4000\AA\ break and she classified this
  object 
  as a galaxy. Variability and the exact position and width of the slit can
  be responsible for this deviation.
\item[RX\,J1353.4+5600] Ca H and K and the G band are marked in
  Fig. \ref{spektren} and a 4000\AA\ break is apparent. These features lead to
  $\rm z = 0.370$. The comparatively strong absorption feature at 5530\AA\
  cannot be explained with this redshift. Since we have taken only one
  spectrum the redshift has a tentative character. 
\item[RX\,J1422.6+5801] Two absorption lines with the right 
   separation to be Ca H and K
  are found in spectra with varying resolutions 
  taken at two
  different telescopes. An absorption line at the redshifted position of
  MgII\,2798 is indicated on the Calar Alto 3.5m spectra and the G band is
  clearly visible on the WHT red spectrum. The high redshift of $\rm z =
  0.638$ together with the high X-ray flux makes this object the most X-ray
  powerful BL\,Lac object known up to now.
\item[RX\,J1456.0+5048] The radio image of this faint object is 
  somewhat
  confusing. The NVSS lists a radio source with 4.4\,mJy 12\arcsec\ distant
  from the optical position and
  another bright one with 220.1\,mJy in 43\arcsec\ distance. Low resolution
  catalogues contain only the bright radio source which would lead to high
  positional deviations and an odd position in the $\alpha_{OX} - \alpha_{RO}$
  diagram. With the low flux RX\,J1456.0+5048 has a typical multifrequency
  spectrum for a X-ray selected BL\,Lac object. The absorption features
  which were used for the redshift determination (Ca H and K and G band) are
  confirmed by another spectrum with the same telescope.
\item[RX\,J1458.4+4832] This object is barely visible on the digitized POSS I
  data. Furthermore the identification of this object is confused by
  another brighter (B=17.7) starlike object near the X-ray position. A
  spectrum taken from this object was classified as 
  of stellar spectral type K.
  For this spectral class the star is too faint to be a plausible optical
  counterpart. Furthermore the NVSS radio position is only consistent with the
  position of the BL\,Lac object. 
\item[1ES\,1533+535] This object has extremely weak absorption features
  hardly above the noise level. The red part of the WHT spectrum shows two
  broad absorption features with a 
  separation consistent with Ca H and K. The blue
  part of the WHT spectrum shows two neighbouring lines with a position
  consistent with the MgII doublet and a redshift of $\rm z = 0.89$. 
  Blended together these lines can also be 
  discerned in the Calar Alto 3.5m spectra. No other line could be
  securely detected at this redshift (in particular the G band). Therefore we
  consider this redshift as very tentative.
  This is the highest redshift in the sample, but we have already argued for 
  weak absorption features in high redshift objects. The main conclusions of
  our paper are not altered if the redshift of this object is considered as
  unknown. This object is also a member of the {\em Einstein} Slew Survey.
\item[RX\,J0832.8+3300] The 4000\AA\ break of this high redshift BL\,Lac object
  is comparatively strong.
\item[RX\,J1237.0+3020] This is another faint object with high 
  redshift.
  Ca H and K and G band are discernible on the red part of the WHT spectrum,
  and the MgII doublet can be found as a blend on the Calar Alto 3.5m spectra.
\end{description}

\end{document}